\documentstyle[12pt]{article}
\begin{document}

\begin{titlepage} 

\begin{flushright}
{UCSB--HEP--94--05\\
DTP/94/12\\
UCD--94--8\\
May 1994\\}
\end{flushright}
 
\vspace{.5cm}

\begin{center}

{\Large\bf Three-jet final states and measuring
the $\gamma\gamma$ width\\[3mm]
of the Higgs at a photon linear collider}

\vspace{.7cm}

        {\bf D.~L.~Borden}\\[1mm]
        {\it Department of Physics, University of California}\\
        {\it Santa Barbara, California 93106, USA}\\[1mm]
        and\\[1mm]
        {\bf V.~A.~Khoze and W.~J.~Stirling}\\[1mm]
        {\it Department of Physics, University of Durham}\\
        {\it Durham DH1 3LE, England}\\[1mm]
        and\\[1mm]
        {\bf J.~Ohnemus}\\[1mm]
        {\it Department of Physics, University of California}\\
        {\it Davis, California 95616, USA}

\end{center}

\vspace{.7cm}

\begin{abstract}
The identification of the intermediate-mass Higgs process
$\gamma\gamma\to H \to b \bar b$ will be one  of the most important
goals of a future photon linear collider. Potentially important
backgrounds from the continuum $\gamma\gamma \to c \bar c ,\/ b \bar b $
leading-order processes can be suppressed by a factor $m_q^2/s$  by
using polarized photon beams in the $J_z = 0$ initial-state
configuration. We  show that the same $m_q^2/s$ suppressions do not necessarily
apply to the  radiative processes $\gamma\gamma \to c \bar c g,\/ b \bar
b g$. These processes can mimic the two-jet topology of the Higgs signal
when two of the three partons are collinear, or when one of the partons
is soft or directed down the beam pipe. We calculate the contribution of
these processes to the two-jet background in the $J_z = 0$ channel. The
largest background is  from the  $\gamma\gamma \to c \bar c g  \to 2\
\mbox{jets}$ process, which yields a cross section in excess of the
Higgs signal.  We investigate the  effect of imposing  additional  event
shape, jet width  and secondary vertex cuts on both signal and
background, and show  that with reasonable detector capabilities it
should be possible to reduce   the background to a manageable level. 
\end{abstract}

\end{titlepage}

\section{Introduction} 

The rapid advance of laser technology makes possible the collision of
high-brightness, high-energy photon beams at future linear colliders
\cite{Ginzburg,Telnov,BBC1} through Compton backscattering 
\cite{Arut,Milburn}.  One
particularly intriguing use of such a photon linear collider is to
measure the two-photon decay width of a Higgs boson once it is
discovered \cite{GH,BBC2}.  The $\gamma\gamma$ width of a Higgs boson is
potentially one of its most important properties. The coupling of the
Higgs to two photons proceeds through loops in which any charged
particle with couplings to the Higgs contributes.  A measurement of the
$\gamma\gamma$ width is thus quite sensitive to new physics at even
higher mass scales \cite{GH}.  Supersymmetric models, technicolor
models, and other extensions of the standard model with more complicated
Higgs sectors all predict two-photon couplings which are, in general,
very different from that of the standard model \cite{HiggsHunt,Konig}.  

In a photon linear collider, the $\gamma\gamma$ partial width of a Higgs
boson, $\Gamma(H \to \gamma\gamma)$, is deduced by measuring the
Higgs production cross section in the reaction $\gamma\gamma \to
H \to X$ where $X$ is the detected final state.  The number of
detected events is proportional to the product $\Gamma(H \to
\gamma\gamma)\, B(H \to X)$ where $B(H \to X)$ is the
branching ratio of the Higgs boson into the detected final state $X$. 
Measuring the production cross section then determines this product.  An
independent measurement of the branching ratio, say at an $e^+e^-$
collider in the process $e^+e^- \to ZH \to ZX$, then
allows a determination of the $\gamma\gamma$ partial width.

For a Higgs boson in the intermediate-mass region,
50~GeV \raisebox{-.6ex}{$\stackrel{<}{\sim}$} 
 $M_H$ \raisebox{-.6ex}{$\stackrel{<}{\sim}$} 150~GeV, 
the dominant decay mode is to $b\bar{b}$.  Measurement of the 
two-photon partial width of the Higgs in this mass region requires
suppressing the continuum $\gamma\gamma \to b\bar{b},c\bar{c}$
background beneath the resonant $\gamma\gamma \to H \to
b\bar{b}$ signal, assuming light quarks can be distinguished from heavy
quarks by vertexing \cite{BBC2}.  The continuum background can be
greatly suppressed by using polarized photon beams. The Higgs signal is
produced by photons in a $J_z=0$ initial state, whereas the continuum
backgrounds are primarily produced by photons in the $J_z=\pm 2$ initial
state, the $J_z=0$ cross section being suppressed for 
large angles by a factor of
$m_q^2/s$ \cite{GH,Isparin}.  

It is important to note that the $m_q^2/s$ suppression of the $J_z=0$
$\gamma\gamma\to q\bar{q}$ cross section is in principle  removed by the
presence of an additional gluon in the final state. It follows that
$\gamma\gamma \to q\bar{q}g$ with $q=b,c$ could be a significant
background for Higgs detection. This process can mimic a two-jet event 
(the dominant signal topology) in two important ways: (i) if two of the
three partons are collinear, for example a fast quark recoiling against
a collinear quark and gluon, or (ii) if one of the three partons is
either quite soft or is directed down the beampipe and is therefore not
tagged as a distinct jet. A particularly interesting example of the
latter is when one of the incoming photons splits into a quark and an
antiquark, one of which carries most of the photon's momentum and
Compton scatters off the other photon,  $q (\bar q)\gamma \to q (\bar q)
g$. Two jets are then identified in the detector, with the third jet
remaining undetected.

In this paper we study the impact of the radiative $q\bar{q}g$
production process on the study of an intermediate-mass Higgs boson at a
photon linear collider. We first describe the calculation of the matrix
element and discuss the various configurations which could be tagged as
two-jet events. In Section 3, we perform a detailed experimental
simulation and compare the resulting background cross sections with
those expected from Higgs production. Finally, our conclusions are
presented in Section 4.

\section{Matrix Elements and Cross Sections}

\subsection{Higgs Production}

For Higgs bosons in the intermediate-mass region, the beam energy spread
of a $\gamma\gamma$ collider is much greater than the total width of the
Higgs boson, and so the number of $H \to b\bar{b}$ events expected
is
\begin{equation}
\label{numHiggs}
N_{H \to b\bar{b}}= 
  \left.\frac{dL_{\gamma\gamma}^{J_z=0}}{dW_{\gamma\gamma}}\right|_{M_H}
  \frac{8\,\pi^2\,\Gamma(H \to \gamma\gamma)\,
        B(H \to b\bar{b})} {M_H^2}
\end{equation}
where $W_{\gamma\gamma}$ is the two-photon invariant mass.
Figure~1 shows the production rate for $\gamma\gamma \to H \to b\bar{b}$
events in the standard model with a typical value of 0.2
$\mbox{fb}^{-1}/\mbox{GeV}$ taken for $dL/dW$.  The width and branching
ratio are taken from Ref.~\cite{HiggsHunt} and a top quark mass of
150~GeV is assumed.

\subsection{Non-radiative background}

The non-radiative ($\gamma\gamma \to b\bar{b},c\bar{c}$)
continuum background cross section is given by
\begin{eqnarray}
\frac{d\sigma(\gamma\gamma \to q\bar{q})}{d\cos\theta} &=&
\frac{12\,\pi\,\alpha^2\,Q_q^4}{s}\frac{\beta}{(1-\beta^2\cos^2\theta)^2}
\\[5pt]
&\phantom{=}& \times 
    \left\{\begin{array}{ll}
    1-\beta^4 & \ \mbox{for}\ J_z=0\\
    \beta^2(1-\cos^2\theta)(2-\beta^2+\beta^2\cos^2\theta) &
                                                  \ \mbox{for}\ J_z=\pm 2
\end{array} \right.
\nonumber
\end{eqnarray}
where $\beta \equiv \sqrt{1-4m_q^2/s}$ is the velocity of the outgoing
quarks, and $m_q$  and $Q_q$ are the mass  and fractional 
electric charge of the quark respectively. The $\gamma\gamma$ center-of-mass
collision energy is $W_{\gamma\gamma}=\sqrt{s}$. Note
the strong $\cos\theta$ dependence of the cross section and that the
$J_z=0$ cross section vanishes, for $|\cos\theta| < 1$, 
 in the high-energy ($\beta \to
1$) limit.  This background can therefore be significantly reduced by
using polarized beams and cutting on $\cos\theta$.

Direct comparison of the continuum background cross sections with the
resonant signal cross section is difficult.  As indicated in
Eq.~(\ref{numHiggs}), the event rate of signal events is proportional to
$dL/dW$ while the event rate for the continuum background is
proportional to the total luminosity; comparing the two requires
choosing a suitable integration range for $W$.  In comparing signal (S)
to background (B) cross sections, we have chosen to normalize the signal
cross sections as if $(dL/dW)_S = (L)_B/(10\mbox{~GeV})$.  This is
equivalent, for the purposes of comparison, to assuming that the
experimental resolution on reconstructing the Higgs mass is 10~GeV.

Figure~2 shows two-photon cross sections for $b\bar{b}$ and $c\bar{c}$
production in polarized collisions and demonstrates the very large
suppression that is possible with polarized photons in the $J_z=0$
state.  A cut of $|\cos\theta| < 0.7$ has been applied.  For comparison,
the Higgs boson signal has been superimposed, with the normalization as
described in the previous paragraph. It is clear that a high degree of
polarization will be crucial in suppressing these continuum backgrounds
below the Higgs signal.

Before discussing the radiative background we comment briefly on the
 origin of the large-angle 
suppression of $q\bar{q}$ production in the
$J_z = 0$ channel as $m_q^2/s \to 0$. Consider the symmetry properties
of the Born amplitude in the $\beta \to 1$ limit.  Because of helicity
conservation at the photon vertices, only amplitudes with opposite
helicities for the quark and antiquark survive. However, the combined
impact of $C$-, $P$- and $T$-invariance, photon Bose statistics, and unitarity,
can be shown to lead to a vanishing amplitude in this limit {\it at
lowest order in perturbation theory}. It follows that all interferences
between the Born and higher-order non-radiative 
diagrams also vanish for the $J_z = 0$
case. In fact for the special case of scattering at angle $\theta = 90^\circ$,
 the vanishing of  {\it all} $J_z = 0$ non-radiative
amplitudes (i.e. not just at leading order) follows simply from rotational 
invariance about the fermion direction and photon Bose statistics.
 For this particular angular configuration the
$T$-invariance argument is redundant.

\subsection{Radiative background}

The non-radiative backgrounds discussed above were considered in
Ref.~\cite{BBC2} with regard to Higgs physics in $\gamma\gamma$
collisions.  With highly polarized beams, such backgrounds are found to
be small and do not hinder the study of an intermediate-mass Higgs
boson at a $\gamma\gamma$ collider.  This raises the question of whether
previously ignored backgrounds could in fact be dominant, or at least
could contribute significantly.

While the lowest order $q\bar{q}$ large-angle cross sections 
are ${\cal O}(\alpha^2/s)$
and ${\cal O}(\alpha^2m_q^2/s^2)$ for $J_z = \pm 2$ and $0$, respectively, the
$q\bar{q}g$ cross sections are ${\cal O}(\alpha^2\alpha_s/s)$ in both cases,
i.e. the $\gamma\gamma \to q\bar{q}g$ cross section is in principle
{\it not} suppressed in
the $J_z=0$ channel at high energies as is the non-radiative cross
section. Furthermore, as we shall see below, there are regions of phase
space where the three-parton final state may be  tagged as a two-jet
event.  In the case of $b\bar{b}g$ and $c\bar{c}g$, the event may have a
vertex structure similar to the non-radiative case, in which case this
process could easily be misidentified as a $b\bar{b}$ final state. In
contrast, the $J_z=\pm 2$ cross section for $\gamma\gamma \to q\bar{q}g$
is simply an ${\cal O}(\alpha_s)$ correction to the much larger
$J_z=\pm 2$ $\gamma\gamma
\to q\bar{q}$ cross section and will not be considered further here.

The full matrix element squared for $\gamma\gamma \to q\bar{q}g$
with massive quarks is too long to write down here, but the matrix
element with massless quarks is particularly simple and contains most of
the important physics.  As a first step, we examine the massless cross
section in detail, reserving the consideration of the massive case until
later.  In all that follows (both massless and massive cross sections)
the following labelling conventions are adopted:
\begin{equation}
\gamma(\lambda_1,k_1) + \gamma(\lambda_2, k_2) \to 
q(p) + \bar{q} (\bar{p}) + g(k) \>,
\end{equation}
where the $\lambda_i$'s are the photon helicities and the $k$'s and
$p$'s are the particle four-momenta.  	

\subsubsection{Massless Quarks}

In the limit of vanishing quark masses, the $J_z=0$ $(\lambda_1 =
\lambda_2)$ matrix element squared for $\gamma\gamma \to q\bar{q}g$ is
given by \cite{GW} 
\begin{equation}
\left| M_{J_z=0}^{} (\gamma\gamma \to q\bar{q}g) \right|^2 
  = 32\,g_s^2\,e^4\,Q_q^4\> 
  \frac{(p\cdot\bar{p}) \left[(p\cdot k)^2+(\bar{p}\cdot k)^2\right]}
  {(p\cdot k_1)(p\cdot k_2)(\bar{p}\cdot k_1)(\bar{p}\cdot k_2)} \>. 
\label{nomass}
\end{equation}
It is instructive to write the cross section in terms of the quark and
antiquark energies.  Note that the final-state parton kinematics are fully
specified by these two energies and three Euler angles which give the
orientation of the final state with respect to the initial state.
Defining  
\begin{equation} 
x \equiv 2\,p_0/\sqrt{s}, \quad \bar{x} \equiv 2\,\bar{p}_0/\sqrt{s},
\quad \cos\theta \equiv p_z/p_0, \quad \cos\bar{\theta} \equiv
\bar{p}_z/\bar{p}_0,
\end{equation} 
the cross section is given by 
\begin{eqnarray}
d\sigma_{J_z=0}^{} (\gamma\gamma \to q\bar{q}g) &=& 
\frac{16\,\alpha_s\,\alpha^2\,Q_q^4}{\pi^2\,s}
\; \frac{(x+\bar{x}-1)\left[(1-x)^2+(1-\bar{x})^2\right]}
{x^2\,\bar{x}^2} \\[5pt]
&\phantom{+}& \times
\frac{dx\,d\bar{x}\,d\alpha\,d\cos\beta\,d\gamma}
     {(1-\cos^2\theta)(1-\cos^2\bar{\theta})} \>, \nonumber
\end{eqnarray}
where $\alpha$, $\beta$, and $\gamma$ are the Euler angles.

Although at the parton level this process results in a three-particle
final state, in practice the event topology following fragmentation and
hadronization may appear to be two-jet-like.  This can occur in two
distinct ways:  two of the three partons may be collinear and so will
appear as a single jet [see Fig.~3(a)]; or one of the partons may be
soft or may be directed down the beampipe and so  not recognized as a
distinct jet [see Fig.~3(b)]. 

In the approximation that the detector covers $4\pi$ of solid angle, the
two- versus three-jet nature of the cross section is independent of the
orientation of the final state and so depends only on $x$ and $\bar{x}$.
The cross section is defined over the Dalitz-plot 
triangle in $x$--$\bar{x}$ space
shown in Fig.~4. In general, the two-jet-like region corresponds to the
periphery of the triangle while the three-jet-like events are confined
to the interior. In the two-jet region, the collinear regime corresponds
to the edges of the triangle while the soft-parton regime corresponds to
the corners. Note that the ($x\approx 1,\bar{x}\approx 1$) corner of the
triangle  is the region of soft gluon
emission.  In this corner the quark and antiquark are energetic and
back-to-back, with the gluon being quite soft.  The $q\bar{q}g$ cross
section is highly suppressed here; in fact, the differential cross
section behaves as $d\sigma/dE_g \sim E_g^3$.\footnote{The physical
origin of this behaviour can be understood by recalling the celebrated Low
expansion \cite{Low} of the matrix element  in powers of
$E_g$, extended to the case
of charged fermions by Burnett and Kroll \cite{Burnett}.}
 This is in marked contrast
to the $J_z = \pm 2$ case, where the cross section exhibits the standard
infrared behaviour $d\sigma/dE_g \sim E_g^{-1}$. In the other corners of
the triangle, ($x\approx 1,\bar{x}\approx 0$) and ($x\approx
0,\bar{x}\approx 1$), it is one of the quarks which is soft. 

Discriminating two- from three-jet topologies on an event-by-event basis
requires specifying a jet-finding algorithm.  A convenient formalism to
use is a clustering formalism, exemplified by the JADE algorithm
\cite{JADE}.  In such a scheme, particle pairs with low invariant mass
are iteratively combined into one particle (by adding their
four-momenta) until no remaining pair has squared invariant mass below
some cutoff.  In general, the cutoff is specified as a fraction of the
total event invariant mass squared and is traditionally called $y_{\rm
cut}$.  Pure $q\bar{q}$ events are efficiently tagged with a $y_{\rm
cut}$ of 0.02--0.03.

\subsubsection{Collinear regime}

If the JADE algorithm is applied at the parton level to the
$\gamma\gamma  \to q\bar{q}g$ process, some simple approximations allow
an analytic expression for the two-jet cross section as a function of
$y_{\rm cut}$.  If $y_{\rm cut} \ll 1$, the region of integration is
confined to the very edges of the $x,\bar{x}$ triangle, where one parton
takes nearly half the event energy and the other two partons are
collinear and recoil against it.  Assuming that all three parton momenta
are nearly collinear allows the cross section to be  integrated
analytically.  Taking the resulting two final-state jets to lie in the
central region of the detector, with $|\cos\theta_{\rm thrust}| <
\cos\theta_0$, the $\gamma\gamma \to q\bar{q}g \to 2\ \mbox{jet}$ cross
section is given by:
\begin{eqnarray}
\label{qqg approx}
\sigma ({q\bar{q}g \to 2\ \mbox{jets}}) &=&
\frac{128\,\alpha_s\,\alpha^2\,Q_q^4}{s}
\> F_1(\cos\theta_0) \> G(y_{\rm cut})\>,\\[5pt]
\nonumber
F_1(z) &=& \frac{1}{4}\ln\!\left( \frac{1+z}{1-z} \right)
+ \frac{z}{2(1-z^2)}\>,\\[5pt]
\nonumber
G(y) &=& 7-\frac{2}{3}\pi^2 + \left[
\frac{2y(y^3+3y^2-11y-9)}{(1+y)^2(1-y)}+4\ln\!\left(\frac{1+y}{1-y}\right) 
\right] \ln\!\left( \frac{2y}{1-y} \right) \\[5pt]
\nonumber
& & +\, \frac{(3y-1)(y^3-5y^2-y+7)}{(1+y)(1-y)^2} + 
4\,\mbox{Li}_2\!\left(\frac{1-y}{1+y}\right) 
- 4\, \mbox{Li}_2\!\left( \frac{2y}{1+y} \right) ,
\end{eqnarray} 
where the dilogarithm function is defined by 
\begin{equation}
\mbox{Li}_2(x) \> = \> -\int_0^x \, {\ln(1-t)\over t} \, dt 
                \>  = \> \sum_{n=1}^{\infty} \, {x^n \over n^2} \>.
\end{equation}
Note that when $y$ is small, $G(y) \sim 2y\ln(1/2y)-y$.  In
particular, for $y_{\rm cut}$ values from 0.01 to 0.1, $G(y_{\rm
cut})$ runs from about 0.07 to 0.3.

Compare Eq.~(\ref{qqg approx}) to the non-radiative $J_z=\pm 2$
cross section, given in the massless limit by 
\begin{eqnarray}
\label{qqsig}
\sigma(q\bar{q}) &=& 
\frac{24\,\pi\,\alpha^2\,Q_q^4}{s} \> F_2(\cos\theta_o) \>,\\[5pt]
\nonumber
F_2(z) &=& \ln\left( \frac{1+z}{1-z} \right) - z \>.
\end{eqnarray}
For $\cos\theta_o = 0.7$, $F_1$ and $F_2$ are 1.120 and 1.035,
respectively.
The $J_z=0$ radiative cross section in this approximation, assuming a
$y_{\rm cut}$ and $\cos\theta_0$ of 0.03 and 0.7 respectively, is
approximately 3\% of the non-radiative $J_z=\pm 2$ cross section.

\subsubsection{Compton regime}

The $\gamma \gamma \to q\bar{q}g \to 2$~jets cross section 
also receives contributions
from configurations where only two of the final state partons appear as
jets in the detector. Inspection of Eq.~(\ref{nomass}) reveals that 
the matrix element squared can become large when one of the final state
quarks is either very soft or is collinear with one of the incoming
photons, e.g. $p\cdot k_1 \to 0$. This important contribution
corresponds to one of the photons splitting into a quark and an
antiquark, one of which undergoes a hard Compton scattering with the
other photon to produce an energetic quark and gluon in the final state
[see Fig.~3(b)]. The extent to which these two jets are back-to-back in
the $\gamma\gamma$ center-of-mass frame (and therefore constitute a
background to $H\to q\bar{q}$) depends on how the momentum is
apportioned between the active and spectator quark in the $\gamma\to q
\bar{q}$ splitting -- the more asymmetric the splitting, the more
back-to-back are the jets. 

To estimate the size of this virtual Compton scattering contribution, we
can use the leading pole approximation \cite{BFK}, i.e.
\begin{eqnarray}
d\sigma (\gamma \gamma \to q\bar{q}g) &\simeq&
d{\cal W}(\gamma \to q\bar{q}) \ 
d{\sigma}(q \gamma \to q g)\vert_{p^* = k_1 - \bar{p}}\>, \\[10pt]
d{\cal W}(\gamma \to q\bar{q}) &=& {\alpha\, Q_q^2 \over 4 \pi^2} \, 
\left[{\bar{x}^2 + (1-\bar{x})^2 \over k_1 \cdot \bar{p}} 
+ {(1-\bar{x})m_q^2 \over (k_1 \cdot \bar{p})^2} \right]
\, {d^3 \bar{p} \over \bar{p}_0} \>,
\label{splitting}
\end{eqnarray}
where $\bar x = 2 \bar{p}_0 / \sqrt{s}$ is the energy 
fraction of the quark which
does not participate in the hard scattering.\footnote{There are of course 
analogous contributions with $q \leftrightarrow \bar q$
and $k_1 \leftrightarrow k_2$.}  For this process to give a two-jet
background, most of the $\gamma \gamma$ scattering energy $\sqrt{s}$
should be deposited in the detector, thus $0 < \bar{x} < \epsilon$ where
$\epsilon$ is a small parameter that will be directly related to the
allowed acollinearity of the two jets in the detector. In particular, if
we use the JADE algorithm to define the two-jet sample then $\epsilon
\sim y_{\rm cut}$.

The transverse momentum integration of the spectator quark gives rise 
to a large logarithm, $\sim \ln(\Delta s/m_q^2)$, where $\Delta s$ is
some fraction of $s$, and so the overall size of this contribution is
roughly
\begin{equation}
\sigma (\gamma \gamma \to q\bar{q}g \to 2\ \mbox{jets})_{\rm Compton}
 \> \simeq \>
{\alpha \, Q_q^2 \over 2\pi} \ {\cal O}(\epsilon) \ 
\ln\!\left({\Delta s\over m_q^2} \right)
\> \sigma(q \gamma \to q g) \>.
\label{logs}
\end{equation}
The form of Eq.~(\ref{splitting}) is correct for unpolarized scattering, 
but in fact there is no particularly strong helicity dependence for
this contribution.  In particular there is no $J_z=0$ suppression in
this case.

Note that the requirement that most of the collision energy should 
be deposited at large angles in  the detector provides a very strong 
suppression of other `resolved photon' contributions, such 
as $\gamma \to gX$ followed by $g \gamma \to q \bar q$. These
processes will therefore not be considered further here.

To summarize, we have identified two potentially important backgrounds
arising from $J_z = 0$ $\gamma\gamma \to q\bar{q}g$ production.
It is instructive to compare these with the leading order background at
the level of couplings and small quantities:
\begin{eqnarray}
\sigma(\gamma\gamma\to q\bar{q})_{\rm LO} &\sim& {\alpha^2 \over s}\;
{\cal O}\left( {m_q^2\over s}\right) \>, \\
\sigma(\gamma\gamma\to q\bar{q}g \to 2\ \mbox{jets})_{\rm collinear} 
&\sim& {\alpha^2 \over s}\;
\alpha_s \; {\cal O}\left( y_{\rm cut} \ln (1/y_{\rm cut}) \right) \>, \\
\sigma(\gamma\gamma\to q\bar{q}g \to 2\ \mbox{jets})_{\rm Compton} 
&\sim& {\alpha^2 \over s}\;
\alpha_s \; {\cal O}(y_{\rm cut}) \, \ln\left( {\Delta s\over m_q^2} \right) \>.
\end{eqnarray}
In Section 3, we will present a detailed experimental study of these
backgrounds. Before doing so, we discuss how the results change when
the quark mass is included in the $q\bar{q}g $ matrix element.

\subsubsection{Massive Quarks}

The matrix element for $\gamma \gamma \to q\bar{q}g$ with
massive quarks and arbitrary initial state photon helicities $\lambda_1,
\lambda_2$ was calculated numerically by both spinor techniques and by
direct computation of the four-component amplitude.  Examples of
calculations using these methods can be found in Refs.~\cite{KLEISS} and
\cite{BARGER}. The numerical results from the two methods agree to
better than 1 part in $10^5$. In both methods the matrix elements were
checked for invariance under changes of the photon and gluon gauge. 
Futhermore, in the massless limit $(m_q \to 0)$ the numerical results
agree with the results from the analytic expression in Eq.~(\ref{nomass}).
Finally, the matrix elements also reproduce the analytic soft-gluon
results for massive quarks.

The main difference between the massive and massless $J_z = 0$ matrix
elements is that the massless matrix element is infrared finite,  
whereas the massive matrix element has an infrared singularity in the
limit $k \to 0$. To illustrate the general features of the $\gamma
\gamma \to q\bar{q}g$ cross sections, we chose $\gamma \gamma$
collisions with center of mass energy $\sqrt{s} = 200$~GeV, $m_b =
4.5$~GeV, and $\alpha=1/137$.  The helicity combination
$\lambda_1=\lambda_2\ (J_z = 0)$ gives the background to $\gamma \gamma
\to H \to b\bar{b}$, so we only consider this helicity combination in
the following discussion. The total $b\bar{b}g$ cross section is calculated
for all three partons in the angular range $|\cos\theta| < 0.7$, and 
the infrared singularity is avoided by imposing a cut $E_g >
1$~GeV.  Figure~5 shows the distribution in gluon energy for
the massive and massless cases.  The very different behavior of the two
cross sections at small $E_g$ is apparent. The effect of the non-zero
$b$-quark mass is also evident near the upper kinematic limit, $E_g =
\sqrt{s}/2$.\footnote{In practice, the behaviour close to the upper kinematic
limit will be strongly modified by higher-order corrections.}
  The two distributions are similar when $m_b \ll E_g \ll
\sqrt{s}$.  The effect of the infrared singularity becomes weaker at
higher collision energy. This is illustrated by the second set of curves 
for $\sqrt{s}=500$~GeV in Fig.~5.

In the massless case the gluon prefers to be hard because, as we have
seen, the matrix element suppresses soft gluons; for $E_g \ll \sqrt{s}$
the cross section behaves as
\begin{equation}
{d\sigma \over dE_g}(\gamma \gamma \to q\bar{q}g, J_z=0,m_q=0) \>\sim\>
  \alpha^2 \, \alpha_s \,
{E_g^3 \over s^3} \;  [ \ldots ] \>.  
\end{equation}

In the massive case, the matrix element has an infrared singularity in
the limit $k\to 0$.  In this limit, the matrix element factorizes into a
`probability of soft gluon emission' times the lowest order matrix
element,
\begin{equation}
\lim_{k \to 0}\ \left| {\cal M}_{\lambda_1,\lambda_2}^{} 
(\gamma \gamma \to q\bar{q}g) \right|^2
\ \to \ {\cal S} (p,\bar{p};k) \>
  \left| {\cal M}_{\lambda_1,\lambda_2}^{} (\gamma \gamma \to q\bar{q}) 
\right|^2 ,
\end{equation}
where
\begin{eqnarray}
{\cal S}(p,\bar{p};k) \!&=&\! g_s^2 \,C_F 
\left[ {2\,p\cdot \bar{p}\over (p \cdot k)
(\bar{p} \cdot k) } - {m_q^2 \over (p \cdot k)^2 } 
- {m_q^2 \over (\bar{p}\cdot k)^2} \right] , \\[5pt]
\!\!\!\! \vert {\cal M}_{\lambda_1,\lambda_2}^{} 
(\gamma \gamma \to q\bar{q}) \vert^2
\!&=&\! {6 e^4 Q_q^4 \over t^2 u^2}  
\left[ 2m_q^2s^2(s-2m_q^2) + (1-\lambda_1\lambda_2)
(t^2+u^2)(tu-2m_q^2s) \right] ,
\end{eqnarray}
and $s=2\,k_1 \cdot k_2$, $t=-2\,k_1 \cdot p$, $u=-2\,k_1 \cdot \bar{p}$.
In the infrared limit and with small quark masses, i.e. $E_g \ll m_q \ll
\sqrt{s}$, the cross section behaves as
\begin{equation}
{d\sigma \over dE_g} (\gamma \gamma \to q\bar{q}g, J_z = 0, m_q \ne 0) 
\> \sim \> \alpha^2 \, \alpha_s \>
{m_q^2 \over s^2\, E_g} \> [ \ldots ] \>.
\end{equation}
In the total cross section, this infrared singularity is cancelled 
by one-loop virtual-gluon corrections to the lowest order $\gamma \gamma
\to q\bar{q}$ process.  The net effect is a finite ${\cal O}
(\alpha_s)$ correction,
\begin{equation}
\sigma_{J_z = 0} (\gamma \gamma \to q\bar{q}g) = 
\sigma_{J_z = 0}^{}     (\gamma \gamma \to q\bar{q} )
\, \left[ 1 + \alpha_s \, C + {\cal O}(\alpha_s^2) \right] \>,
\end{equation}
with $C$ a known coefficient, see for example \cite{NLO}.
To avoid spurious large contributions from the soft-gluon region, in what
follows we will impose a cut $E_g > E_{\rm min} = \sqrt{s}/10$. None of
our results depend sensitively on this parameter.
We should mention also that the same infrared problems are encountered
in the next-to-leading order Higgs decay process $H\to b \bar b g$,
where the addition of  virtual-gluon corrections lead to a finite
${\cal O}(\alpha_s)$ correction to the leading-order decay width.

\section{Experimental Considerations}

The $J_z = 0, \gamma \gamma \to q\bar{q}g$ cross section, even for small 
values of $y_{\rm cut}$, is a few per cent of the $J_z = \pm 2, \gamma
\gamma \to q\bar{q}$ cross section.  This cross section for bottom and
charm quarks, in the approximation of Eq.~(\ref{qqg approx}) with
$y_{\rm cut} = 0.02$, along with the non-radiative backgrounds is shown
in Fig.~6.

In a photon linear collider, it is possible to achieve a $\frac{J_z=0}
{J_z=\pm 2}$ ratio of 20--50, so in order to bring the {\it rates} for
the radiative processes down well below that of the non-radiative
processes, it is necessary to find cuts which further reduce the
radiative backgrounds by a factor of about 5--10, without seriously
degrading the $H \to b\bar{b}$ signal.  In order to explore whether 
this is possible, we have employed a Monte Carlo integration of the
radiative cross section with massive quarks which includes fragmentation
and hadronization (via JETSET 6.3 \cite{JETSET}) and a simple detector
simulation.  The detector simulation is a gaussian smearing of the final
state four-momenta by resolutions typical of detectors considered for a
Next Linear Collider, such as the JLC detector \cite{JLC}.  Vertexing,
tracking, and calorimetry are all simulated, but particle identification
is not.

We assume a $\gamma\gamma$ invariant mass of 100 GeV and require that
the thrust axis satisfy $|\cos\theta|<0.7$.  Imposing a $y_{\rm cut}$
of 0.02 to define two-jet events, the values of the relevant cross
sections are as follows:
\begin{eqnarray}
\label{summary}
\sigma_{J_z=0}(\gamma\gamma \to H \to b\bar{b} \to 2\ \mbox{jets}) &=&
0.86\ {\rm pb}\>, \nonumber \\[5pt]
\nonumber
\sigma_{J_z=\pm 2}(\gamma\gamma \to b\bar{b} \to 2\ \mbox{jets}) &=&
2.21\ {\rm pb}\>, \nonumber \\[5pt]
\sigma_{J_z=\pm 2}(\gamma\gamma \to c\bar{c} \to 2\ \mbox{jets}) &=&
35.6\ {\rm pb}\>, \\[5pt]
\sigma_{J_z=0}(\gamma\gamma \to b\bar{b}g \to 2\ \mbox{jets}) &=& 
0.035\ {\rm pb}\>, \nonumber \\[5pt]
\sigma_{J_z=0}(\gamma\gamma \to c\bar{c}g \to 2\ \mbox{jets}) &=& 
0.87\ {\rm pb}\>.\nonumber
\end{eqnarray} 

\subsection{Event Shape and Jet Width Cuts}

Although a $y_{\rm cut}$ of 0.02 tends to select very two-jet-like
events, the $q\bar{q}g$ events still tend to be more spherical than
the $q\bar{q}$ events, as shown in Fig.~7(a).  A cut on event sphericity
then further reduces the radiative cross section without greatly
diminishing the $q\bar{q}$ rate.  The efficiency as a function of
sphericity cut is shown in Fig.~7(b). 

Recall that the dominant contribution to the radiative cross section
comes from the virtual Compton configuration, in which the final state
can be described as a hard quark back-to-back with a gluon and a soft quark
nearly at rest.  As gluon jets tend to be broader than quark jets, a cut
on the width of the final state jets preferentially cuts the radiative
final state over the non-radiative final state.  Specifically, we cut on
the opening half-angle of the cone that contains 90\% of the jet energy. 
Fig.~8(a) shows the distribution of this angle for both radiative and
non-radiative processes.  The efficiency as a function of
$\theta_{90\%}$ is plotted in Fig.~8(b). 

Choosing a sphericity cut of 0.02 and a $\theta_{90\%}$ cut of
$20^\circ$ results in the following efficiencies:
\begin{eqnarray}
\epsilon (b\bar{b})=91.2\%\>, &\quad& \epsilon (b\bar{b}g)=15.1\%\>, 
\nonumber\\[5pt]
\epsilon (c\bar{c})=94.4\%\>, &\quad& \epsilon (c\bar{c}g)=34.0\%\>.
\end{eqnarray}

\subsection{Vertex Cuts}

One might think that since the $b\bar{b}g$ final state---in the dominant
kinematic configuration---contains one fast and one slow $b$ quark that
the vertex structure might differ greatly from a $b\bar{b}$ final state
with two fast quarks.  In practice, it turns out that the vertex
structure is 
%not that dissimilar between the two.  
similar for these two final states.
The $B$ hadrons from
the slow $b$ quark tend to be `pulled' into the gluon jet and do give
rise to displaced vertices in the gluon jet.  Vertexing therefore is not
a powerful discriminant between the radiative and non-radiative
processes.
Such a displacement of $B$ hadrons towards the gluon side is a 
consequence of the 
well-known `string' \cite{STRING} or `drag' \cite{DRAG} effect. 
It reflects the fact that particle production is governed by the collective
action of  color-connected partons, in this case the spectator quark and the 
outgoing gluon. Note also that because of the difference in the color
topology of the underlying subprocesses, the structure  of particle flow
in the $q \bar q$ and Compton-regime $q \bar q g$ events  has some distinct
differences (see for example Ref.~\cite{BOOK})
  which, in principle, could be exploited to further discriminate
between these types of events.

Distinguishing charm from bottom, however, will rely crucially on
vertexing, both for the radiative and non-radiative processes.  Although
it is not our intention here to exhaustively examine this issue, the
factor of 16 amplification of the charm cross sections over the bottom
cross sections necessitates at least some discussion.

As $B$ hadrons are long-lived, they tend to travel a finite distance
before decaying, so that their decay products form displaced vertices
which are measureable with modern vertex detectors.  The same is true of
charmed hadrons, but they tend to travel less far than $B$'s and have
fewer tracks with displaced vertices.  Vertexing is therefore a very
useful tool both in separating light $(u,d,s)$ from heavy $(c,b)$ quark
jets and in separating $b$'s from $c$'s.  

Rather than reconstructing each decay vertex from the charged tracks in
an event, it is sufficient to find the impact parameter (distance of
closest approach, either in 3 dimensions or in the $x$-$y$ plane) for 
each track.  Modern vertex detectors are capable of impact parameter
resolutions of $\sim$~30~$\mu$m.  Requiring each event to have, say, 4
or 5 tracks with high $(>4\sigma)$ impact parameter (not including
tracks which form $K_S$'s or $\Lambda$'s) results in the following
efficiencies.

\begin{center}
\begin{tabular}{c|cccc}
            & \multicolumn{2}{c}{2-D} & \multicolumn{2}{c}{3-D} \\ 
process     & 4 tracks & 5 tracks     & 4 tracks & 5 tracks     \\ \hline
$b\bar{b}$  &   57\%   &   37\%       &   77\%   &   61\%       \\
$b\bar{b}g$ &   54\%   &   37\%       &   72\%   &   59\%       \\
$c\bar{c}$  &   3.7\%  &   0.8\%      &   5.9\%  &   1.6\%      \\
$c\bar{c}g$ &   4.4\%  &   1.0\%      &   7.4\%  &   2.0\%      \\
\end{tabular}
\end{center}

Applying a sphericity cut of 0.02, a jet width cut of $20^\circ$, and
requiring 5 tracks with high 3-D impact parameter then results in the
following production cross sections:
\begin{eqnarray}
\label{summary2}
\sigma_{J_z=0}(\gamma\gamma \to H \to b\bar{b} \to 2\ \mbox{jets}) &=&
0.48\ {\rm pb}\>, \nonumber \\[5pt]
\nonumber
\sigma_{J_z=\pm 2}(\gamma\gamma \to b\bar{b} \to 2\ \mbox{jets}) &=&
1.2\ {\rm pb}\>, \nonumber \\[5pt]
\sigma_{J_z=\pm 2}(\gamma\gamma \to c\bar{c} \to 2\ \mbox{jets}) &=&
0.54\ {\rm pb}\>, \\[5pt]
\sigma_{J_z=0}(\gamma\gamma \to b\bar{b}g \to 2\ \mbox{jets}) &=& 
0.0031\ {\rm pb}\>, \nonumber \\[5pt]
\sigma_{J_z=0}(\gamma\gamma \to c\bar{c}g \to 2\ \mbox{jets}) &=& 
0.0059\ {\rm pb}\>.\nonumber
\end{eqnarray} 

\section{Summary and Conclusions}

The identification of the intermediate-mass Higgs boson via the process
$\gamma\gamma\to H \to b \bar b$ will be one  of the most important
goals of a future photon linear collider. Potentially important
backgrounds from the continuum $\gamma\gamma \to c \bar c,\/ b \bar b$ 
leading-order processes can be suppressed by a factor $m_q^2/s$  by
using polarized photon beams in the $J_z = 0$ initial-state
configuration. In this paper we have pointed out that the same $m_q^2/s$
suppressions do not apply to the  radiative processes $\gamma\gamma \to
c \bar c g,\/ b \bar b g$. These processes can mimic the two-jet
topology of the Higgs signal when two of the three partons are
collinear, or when one of the partons is soft or directed down the beam
pipe. Our detailed numerical calculations of the various two-jet cross
sections, summarized in  Eq.~(\ref{summary}), show that these radiative
processes do indeed provide the dominant background in the $J_z = 0$
channel.  Particularly  problematic is the $c \bar c g$ background
which, because of the quark charge, is much larger than the
corresponding $b$-quark process. For our choice of  kinematic cuts and
for a $\gamma\gamma$ collision energy of 100~GeV, the $\gamma \gamma \to
c \bar c g  \to 2\ \mbox{jets}$ background is comparable to
 the Higgs signal. In order to try to reduce this background
further, we have studied the effect of additional event shape, jet width
and vertex cuts. The results, described in Section~3 and summarized in
Eq.~(\ref{summary2}), indicate that further improvements in the signal
to background ratio can indeed be achieved. In particular, a modern
vertex detector should be capable of achieving the necessary rejection
of $c$-quark events while remaining reasonably efficient for the signal
$b$-quark events.

\section*{Acknowledgements}

We would like to acknowledge valuable discussions with Gary Greenbaum,
Mike Strauss, and Mike Hildreth.  This work was supported in part by
the United Kingdom Science and Engineering Research Council, 
by the United States Department of Energy under Grants Nos. DOE-FG-91ER40618
and DE-FG03-91ER40674, and by Texas National Research Laboratory Grant
No. RGFY93-330.

\section*{\large \bf Figures}
 
\begin{enumerate}
\item Cross section for $\gamma \gamma \to H \to b\bar{b}$ as a function
of the Higgs mass. A value of 0.2 $\mbox{fb}^{-1}/\mbox{GeV}$ has been
taken for $dL/dW$.
\item Cross sections for $\gamma \gamma \to b\bar{b}$ and $\gamma \gamma
\to c\bar{c}$ in polarized collisions. A cut of $|\cos\theta| < 0.7$ has
been applied.  For comparison, the Higgs boson signal has been
superimposed, with the normalization as described in the text.
\item Examples of how the $q\bar{q}g$ final state can appear as two
jets: a) two partons are collinear, b) one of the partons is soft or
directed down the beam pipe.  The solid and wavy lines represent
quarks and gluons, respectively.
\item Diagram of $x$--$\bar{x}$ phase space showing the two- and three-jet 
event regions.
\item Distribution of the gluon energy in the process $\gamma \gamma
\to b\bar{b}g$ for massless and massive ($m_b = 4.5$~GeV) $b$-quarks
for center-of-mass energies $\sqrt{s} = 200$ and 500~GeV.
\item Cross sections for $\gamma \gamma \to b\bar{b},c\bar{c}$ and
$\gamma \gamma \to b\bar{b}g,c\bar{c}g$.  The approximaton of 
Eq.~(\ref{qqg approx}) is used for the radiative cross sections.
\item a) Sphericity distribution of $b\bar{b}$ and $b\bar{b}g$ events. 
b) Efficiency as a function of sphericity cut for $b\bar{b}$ and
$b\bar{b}g$ events.
\item a) Distribution of opening half-angles of cones containing 90\% of
the jet energy for $b\bar{b}$ and $b\bar{b}g$ events; a sphericity cut
of 0.02 is included.  b) Efficiency as a function of cut on $\theta_{90\%}$
for $b\bar{b}$ and $b\bar{b}g$ events.
\end{enumerate}

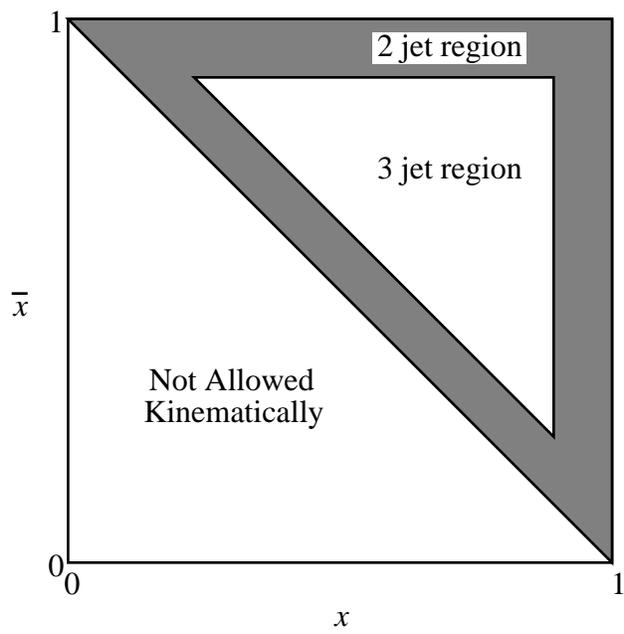
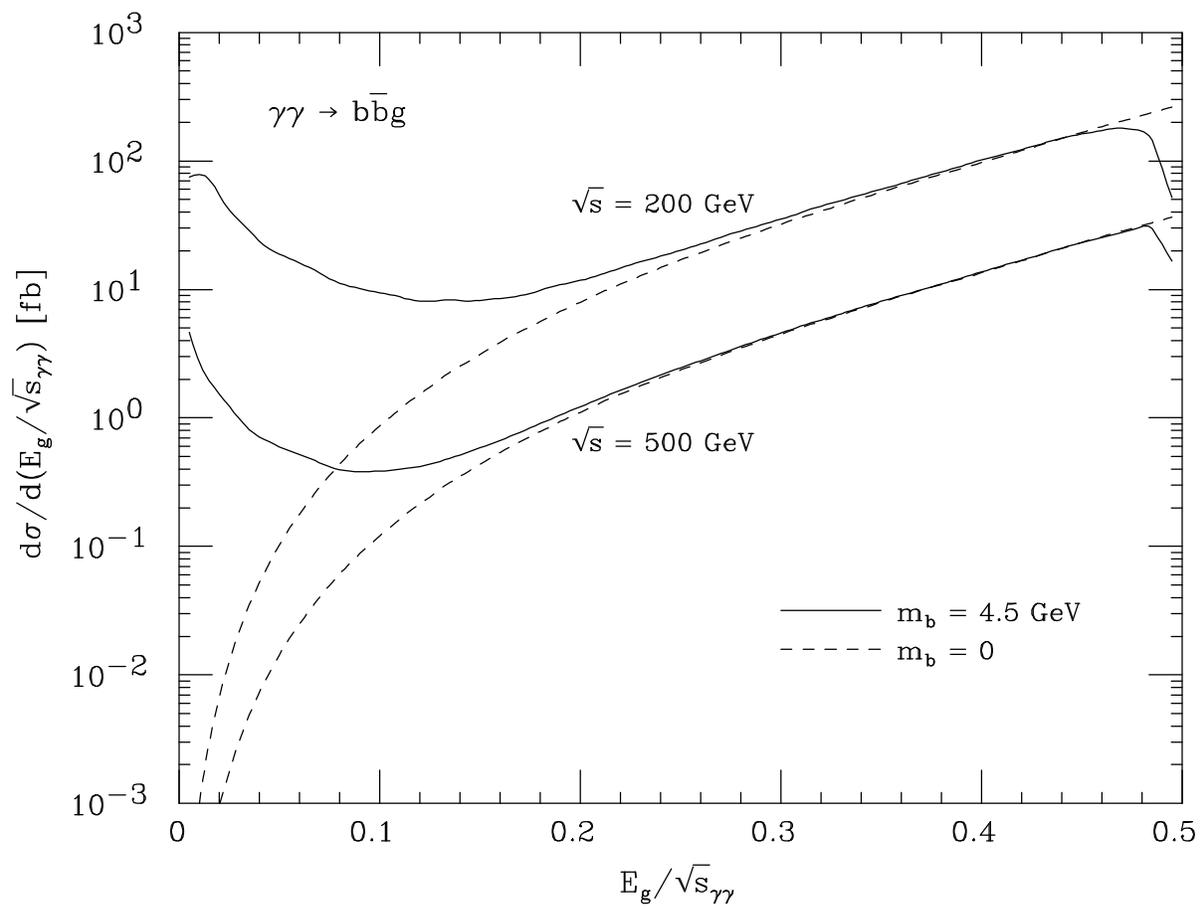
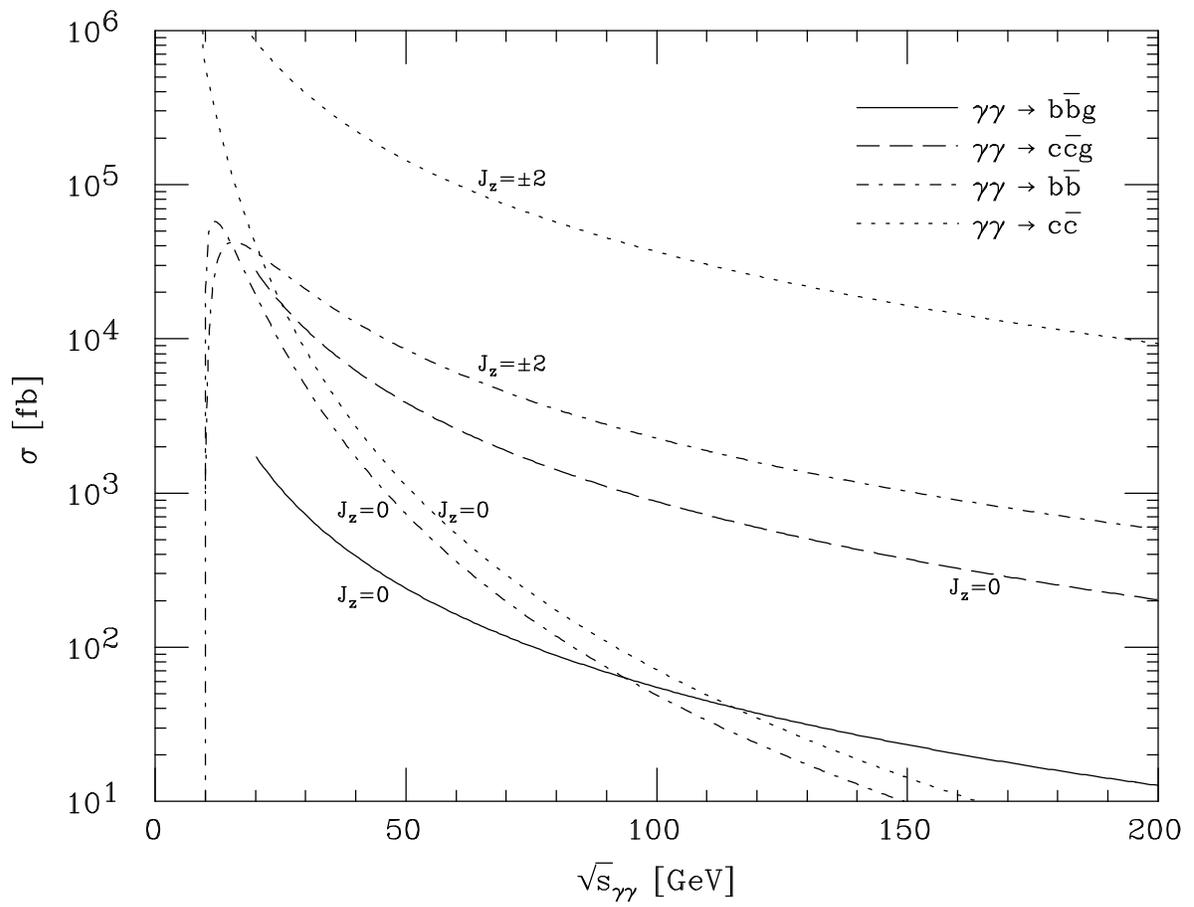
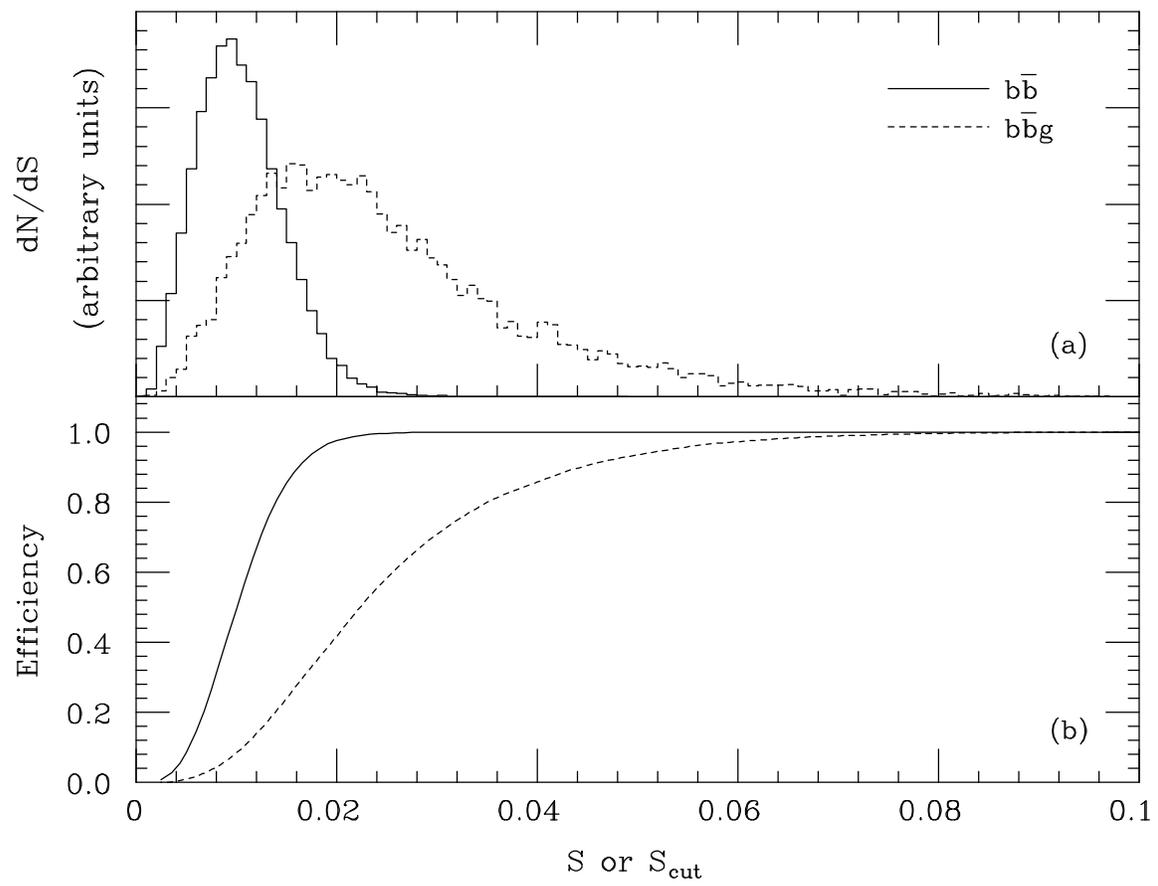
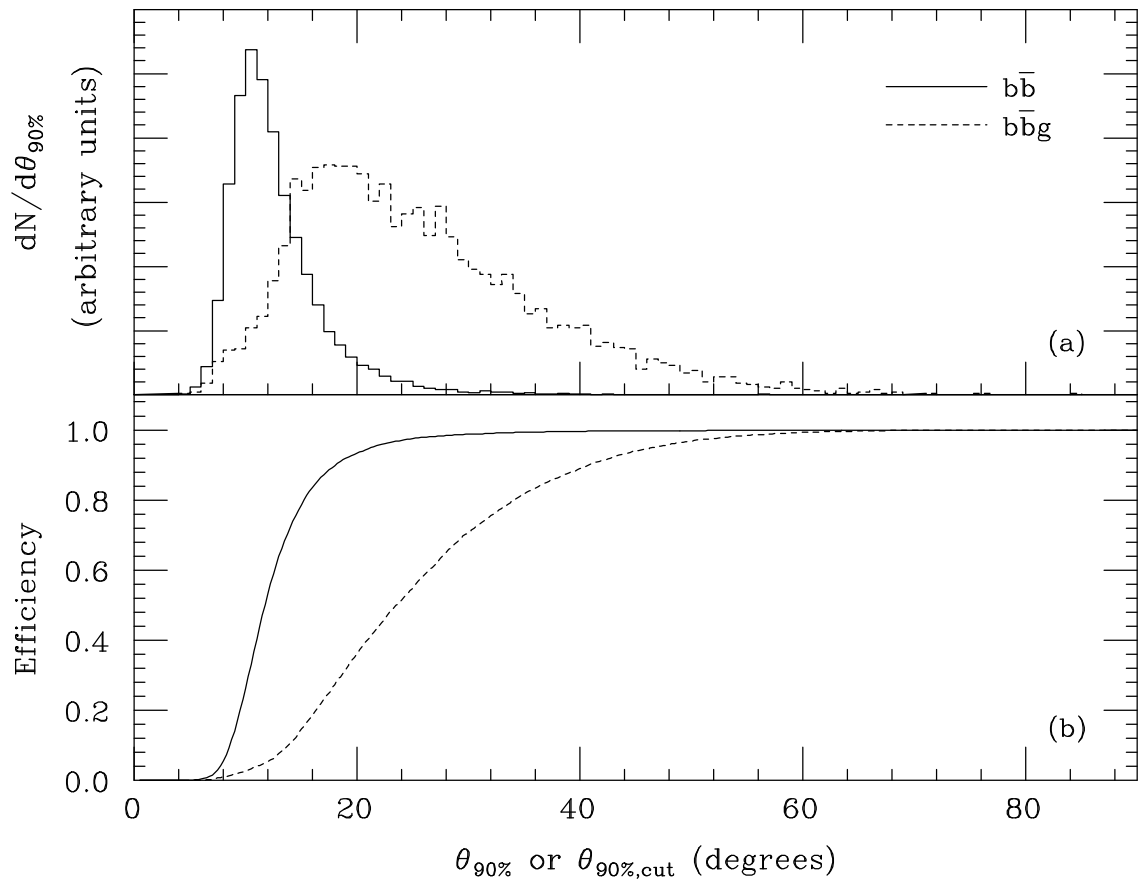
\end{document}